\documentstyle[sprocl]{article}
\def\be{\begin{equation}}
\def\ee{\end{equation}}
\def\bea{\begin{eqnarray}}
\def\eea{\end{eqnarray}}

\begin{document}

\title{A SIMULTANEOUS SOLUTION TO BARYOGENESIS AND DARK MATTER PROBLEMS
\footnote{Talk presented at the International Workshop on Future Prospects of
Baryon Instability Search in p-Decay and $n-{\bar n}$ Oscillation Experiments,
Oak Ridge, Tennessee, March 28-30, 1996, and \\ at the Workshop 'Aspects of Dark
Matter in Astro- and Particle Physics', Heidelberg, Germany, September 16-20,
1996. } }

\author{Vadim A. Kuzmin}

\address{Institute for Nuclear Research of Russian Academy of Sciences,\\
60th October Anniversary Prosp. 7a, Moscow 117312, Russia}

\address{
\begin{center} and \\ \end{center} 
Max-Planck-Institut fuer Physik,\\
Foehringer Ring 6, \\ 
80805 Muenchen, Germany}

\maketitle
\abstracts{
A new concept of generation of the cosmological baryon excess
along with the cold dark matter (CDM) in the Universe is proposed and
corresponding scenarios are outlined. Possible realizations of the idea 
in the framework of supersymmetric models are considered
and constraints (predictions) on masses of sparticles compatible
with the viability of the scenario are derived. Multiple predictions might be 
extracted from the concept. In particular, we predict 
a quite natural existence of a charge asymmetric component of CDM. 
In particular, a $\sim 10^{-2}$ part of CDM might exist in the form of 
electrically 
charged relic particles with masses $m \simeq 1$ TeV. They are negatively 
charged and are dressed by protons. This conjecture provokes a rich 
field of immediate search for these particles.  The charge 
symmetric component of CDM might 
be represented by very light, $m \approx 2$ GeV, very weakly interacting 
particles like right-handed sneutrinos
, so recoils expected are rare and have quite small energies, $E_{recoil} \sim 
1$ KeV. This leads by the way to prediction of long-living sparticles.      
Some new  experimental 
proposals for non-traditional search of cold dark matter particles 
are mentioned. 
}

\section{Introduction.}

Starting with the papers by Sakharov \cite{sakh} and Kuzmin \cite{kuz1} 
where the principal
ways of solving the problem of the baryon asymmetry of the Universe
(BAU) were outlined there was a long list of various attempts of 
elaboration of the main concepts, most convincing in the framework of Grand 
Unified Theories \cite{ign} which naturally provide all the necessary 
conditions for the creation of charge asymmetric state of the matter in 
the Universe starting with the {\it symmetric} one at high temperature. 
This is a beautiful concept, indeed. And, indeed everything seemed to be 
O.K. with the origin of the baryon asymmetry of the Universe in the 
framework of Grand Unified Theories until 1985.  However, after the discovery
was made in 1985 in the paper by Kuzmin, Rubakov and Shaposhnikov \cite {krs} 
that electroweak sphaleron-induced baryon
and lepton number non-conserving transitions might have been not
suppressed in the $SU(2) \times U(1)$ unbroken phase at high
temperatures $T \geq T_{EW} \sim M_{W}$, the GUT based realizations
of the scenario of the BAU
generation  were re-examined in view of this potentially dangerous
washing-out the baryon excess phenomenon and ideas were proposed of
just exploration of sphaleron-mediated transitions for generation of the BAU.
Of particular interest are mechanisms of sphaleron re-processing of
a previously generated lepton number excess considered by Fukugita and 
Yanagida \cite {fuku} and by Langacker {\it et al} \cite{lan} exploring 
the see-saw mechanism of effective lepton number non-conservation. 
Efforts of generation of the BAU within the framework of the
Standard Model (SM) started with the paper by Shaposhnikov \cite{shap} 
are being made as well. Hopefully, these efforts will result in a 
plausible explanation of the cosmological baryon excess. However, at 
present it seems quite problematic to solve the problem within the framework 
of the minimal Standard Model. 

And by the way there is yet another problem which was put
under consideration after observation of presence of dark matter
in the Universe, just the problem of its nature as well as the
origin. There is no room, I mean, no elementary particle candidate in the 
particle spectrum of the Standard Model which may serve as a candidate for the
Cold Dark Matter in the Universe. The axion is the only exception. This 
is definitely still a good candidate. 

It seems  being taken at present (see, e.g. the paper by 
Primack \cite {prim}) that it is just 
the cold dark matter rather than the hot one which
populates the Universe predominantly,
$\Omega_{CDM}h_0^2 \sim 0.7$, the most popular version of dark matter 
content being given by the mixed model, Cold Dark Matter plus Hot Dark Matter, 
something like $\Omega_{CDM} \sim 0.7$, $\Omega_{HDM} \sim 0.2$.  

It is our impression that after all one has to extend the particle content 
beyond the Standard Model in order to find solution to both these   
problems, the BAU and CDM. 

There was already a
number of papers devoted to a combined solution of
both the problem of the BAU and the CDM altogether
(see, e.g. the papers by Barr {\it et al} \cite {barr}, Kaplan {\it et al} 
\cite {kap}, Kuzmin {\it et al} \cite {kuz3}), etc. We would like to 
take part in the race, too, and again.  

\section{Electroweak Sphalerons and Anomalous Fermion Number Non-Conservation.}

In this Section we would like to remind shortly some properties of 
electroweak sphalerons and their role in fast anomalous baryon and 
lepton number non-conservation 
at high temperatures. As one will see electroweak sphalerons are by 
themselves the very powerful tool for a solution of cosmological 
problems rather than destruction of nice constructions. 

The crucial points for the anomalous fermion number non-conservation in the 
electroweak theory with the gauge symmetry $SU(2) \times U(1)$ are :

1. The anomaly in the fermionic currents discovered by Adler, Bell and 
Jackiw \cite {abj}
\begin{equation}
\partial_{\mu}J^{B}_{\mu}=\partial_{mu}J^{L}_{\mu}=
{\frac {n_{f}}{32{\pi}^{2}}} 
(-g^{2}F^{a}_{\mu \nu}{\tilde{F}^{a}_{\mu \nu}}+g^{2}F^{0}_{\mu \nu}{\tilde 
{F}^{0}_{\mu \nu}}),  
\label{eq:J}
\end{equation}
\noindent
where $J^{B}_{\mu}$ and $J^{L}_{\mu}$ are the baryon and lepton currents, 
respectively, $F^{a}_{\mu\nu}$ is the $SU(2)$ field strength and $n_f$ is the 
number of fermionic generations, which at the moment is known to be 
$n_{f} \geq 3$.

2. The nontrivial vacuum structure in non-Abelian gauge theories observed 
by Christ, Dashen and Jackiw \cite{cdj}. 

Topologically distinct vacua are separated by the potential barrier of the 
minimal height $E_{sph} = 2M_{W}/{\alpha_{W}B(\lambda/{\alpha_{W}})}=8-14$ 
TeV for $\lambda$ varying from $0$ to $\inf$ \cite{kman} ($\lambda$ is the 
Higgs self-coupling constant, $\alpha_{W} \sim (1/30)$ is the $SU(2)$ fine 
structure constant). The label (sph) refers to the sphaleron, i.e. the static 
unstable solution to the classical equations of motion found by Klinkhamer and 
Manton \cite{kman}. This configuration 
belongs to the minimal energy path from one vacuum to the other. 

The selection rules for the anomalous processes are :
\begin{equation}
\Delta n_f = 3n_f , \Delta n_l =n_f , \Delta B =\Delta L =n_f .
\label{eq:rul}
\end{equation}

If bosonic configuration changes from one vacuum configuration to another 
one, there always takes place the creation of a net number of fermions 
(or antifermions !) proportional to the change of the Chern-Simons number 
 \cite{christ}.

In the case of zero temperature, low fermionic densities and low energies of 
colliding particles, the initial state of the system as well as the final
state are close to the vacuum configurations. So, in order to provide the 
fermion number non-conservation the system has to tunnel through the energy 
barrier. This process might be described by instantons( see the paper by 
Belavin {\it et al} \cite{bel} and is 
strongly suppressed by the semiclassical exponent as was first shown by 
't Hooft  \cite{tho}, $\exp(-2\pi /{\alpha_{W}})$. 
 
At nonzero temperature, the system experiences thermal fluctuations. Due to 
the equipartition distribution, every degree of freedom is excited and the 
average energy stored in it is of order of temperature. In particular, 
the sphaleron mode is excited as well. 

If the energy of excitation is greater than the potential barrier height, 
then the system travel ${\it classically}$ from the vicinity of one 
topological vacuum to the other. The rate of these transitions leading to 
fermionic number non-conservation is proportional to the Boltzmann 
exponent $\exp(-E_{sph}(T)/T)$ determining the density of negative mode 
excitations with energies higher than the barrier energy \cite{krs} .
Here $E_{sph}(T)=2M_{W}(T)/{\alpha_{W}B(\lambda/\alpha_{W}})$ is the 
effective sphaleron mass accounting for the temperature dependence of the 
Higgs vacuum expectation value, $M_{W}^{2}(T)=M_{W}^{2}(1-T^{2}/T_{c}^{2})$ 
at $T<T_{c}$, where $T_c$ is the temperature of the electroweak phase 
transition as conjectured by Kirzhnits and Kirzhnits and Linde \cite{kir}. 
The calculations of the prefactor by Arnold and McLerran and Shaposhnikov 
\cite{armc,shap} give for the rate 
of the topological transitions per unit volume per unit time 
\begin{equation}
\Gamma = \frac{T^4 \omega_{-}}{M_{W}(T)}(\frac{\alpha_{W}}{4\pi})^{4}N_{tr}
N_{rot}(\frac{2M_{W}(T)}{\alpha_{W}T})^{7}\kappa \exp(-\frac{E_{sph}(T)}{T}),
\label{eq:rate}
\end{equation}
\noindent
where the factors $N_{tr} \sim 26$, $N_{rot} \simeq 5$ are due to the zero 
modes normalizations \cite{armc}, $\kappa \sim 1$ is the determinant of
nonzero modes around the sphaleron and $\omega_{-} \sim M_{W}(T)$ is the 
magnitude of the sphaleron negative mode. At $T<M_{W}$ quantum tunneling 
is more efficient than the classical transitions while for $T>E_{sph}$ 
the saddle point approximation for the rate is not applicable. Moreover, 
at temperatures greater than the critical temperature $T_c$ the $SU(2)$ 
symmetry is restored, the vacuum expectation value of the Higgs field 
is zero and the sphaleron saddle point solution does not exist anymore. 

It is quite clear, however, that the rate of topological transitions changing 
fermion ( baryon and lepton) number is not suppressed by any exponent in the
temperature range $T>T_{c}$ due to absence of the energy barrier between 
topologically different vacua. 

With the use of scaling arguments it may be shown \cite{shap,armc} that 
\begin{equation}
\Gamma =A(\alpha_{W} T)^4 
\label{eq:rate3}          
\end{equation}
\noindent 
where $A$ is some factor which cannot be found by semiclassical methods. 
The real time numerical simulations give the value $A \approx 0.1-1.0$
\cite{}.

At temperatures larger than the critical one, $T>T_c$, the rate Eq. ~\ref
{eq:rate} of the anomalous processes with baryon number non-conservation 
greatly exceeds the rate of the Universe expansion rate, $t_{U}$,  
\begin{equation}
t_{U}^{-1} =T^{2}/M_{0} , ~~M_{0}=M_{Pl}/1.66N_{eff}^{\frac{1}{2}}
\label{eq:Univ} 
\end{equation}
where $N_{eff} \sim 100$ in the case of Standard Model is the effective 
number of massless degrees of freedom at this temperature. 

Therefore, the anomalous reactions violating baryon and lepton numbers 
are in thermal equilibrium till the time of the electroweak phase 
transition. After the phase transition the Higgs field develops the 
non-vanishing vacuum expectation value and as a result the rate of baryon 
and lepton number violating processes decreases rapidly due to the Boltzmann 
exponential suppression. 

Summarizing, one may say that at high temperatures, $T>T_c$, there are 
very fast transitions ( we shall call them 'sphaleron-mediated' transitions) 
which result exactly in the following 
\begin{equation}
|vacuum> \rightarrow 9 (quarks) + 3 (leptons) 
\label{eq:vac1}
\end{equation}
and
\begin{equation}
|vacuum> \rightarrow 9 (antiquarks) + 3 (antileptons) .
\label{eq:vac2}
\end{equation}   

These are the processes which re-process any $B$- or $L$-excess in the 
normal Standard Model fermionic sector distributing it correspondingly 
between quarks and leptons. The net $B-L$ remains, of course, intact 
since in the Standard Model $B-L$ is conserved both perturbatively and 
non-perturbatively. Sphalerons do respect $B-L$ conservation as well.

Now we are going to describe a possible scheme of the simultaneous 
genesis of the cosmological 
baryon excess and the cold dark matter in the Universe.

\section{The Mechanism.}

Let there exist in nature some new kind of baryon (lepton) number
bearing particles (called in what follows $R_{q}$ and ($R_{l}$)).
interacting with the Standard Model quarks and leptons.
We are not going to assume {\it {a priori}} that there exist any new 
interactions
in addition to the standard $SU(3)\times SU(2)_L \times U(1)$ ones, i.e.
we extend just the particle content of the SM.

As Abdus Salam said,\\

We Have to be Economical in Principles

Rather Than in Structures.\\

The crucial requirement to these new baryon (lepton) number bearing
R-particles is that unlike normal (left-handed)
fermions they are to be 'EW-sphaleron-blind', i.e. the  R-currents are
to be EW non-anomalous.
This means that R-particles should be either bosons (case 1) or
${SU(2)}_{L}$-singlet fermions with the ineffective enough, at least
at some temperature, chirality equilibration rate (case 2). At present,
let us restrict ourselves by the case 1, the R-particles
being just bosons (like sfermions in supersymmetric models).

Now our basic idea is as follows.

Let the state of cosmological plasma with $B-L \equiv (B-L)_{init} \neq 0$
in the normal SM sector and $B-L = -(B-L)_{init}$ in the R-sector be
somehow created at some temperature $T^{\ast}>T_{EW}\sim 10^{2}$ GeV,
$T_{EW}$ being the effective temperature of switching-off un-suppressed
electroweak transitions violating baryon, lepton and fermion numbers (see
Fig. 1).

In other words, let there occur in the Universe an ${\it asymmetrization}$ of
plasma with respect to $B-L$ distribution between the normal SM
fermionic sector of Standard Model and the new sector R.
For definiteness, let the normal left-handed fermionic
sector acquire some $(B-L)_{init}<0$ and the R-sector $(B-L)_{init}>0$,
the overall $B-L$ of plasma being exactly preserved. If such a phenomenon 
took  place then this might be all one needs to understand the origin of 
the baryon excess ${\it and}$ the dark matter in the Universe.

\begin{center}
\setlength{\unitlength}{1mm}
%\begin{center}
\begin{picture}(140,70)(0,0)
\put(40,20){\vector(-1,0){35}}
\put(40,20){\vector(1,0){70}}
\put(23,20){\line(0,1){5}}
\put(23,28){\line(0,1){5}}
\put(23,36){\line(0,1){5}}
\put(23,44){\line(0,1){5}}
\put(23,52){\line(0,1){5}}

\put(67,20){\line(0,1){5}}
\put(67,28){\line(0,1){5}}
\put(67,36){\line(0,1){5}}
\put(67,44){\line(0,1){5}}
\put(67,52){\line(0,1){5}}

\put(08,14){$T$}
\put(107,14){$t$}

\put(23,14){$T^{\ast}$}
\put(67,14){$T_{EW}$}

\put(27,51){$SPHALERONS$}
\put(65,45){\vector(-1,0){10}}
\put(27,39){$OPERATING$}
\put(65,33){\vector(-1,0){10}}

\put(72,51){$SPHALERONS$}
\put(69,45){\vector(1,0){10}}
\put(72,39){$FROZEN-OUT$}
\put(69,33){\vector(1,0){10}}
\end{picture}
\end{center}
\begin{center}

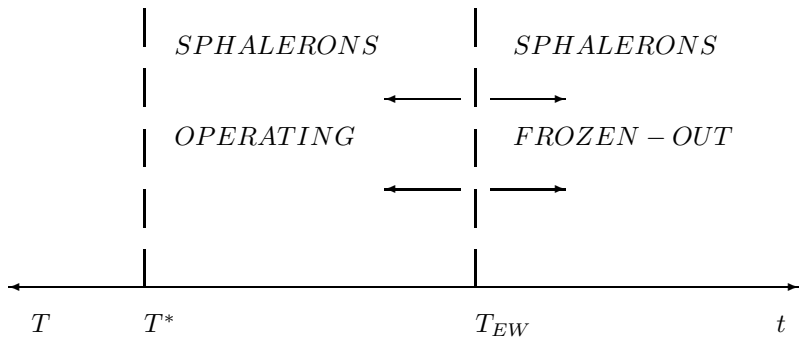
Fig. 1. A schematic picture of a temperature evolution of the $B(L)$
distribution in cosmological plasma. At $T\simeq T^{\ast}$ plasma is symmetric
with respect to $B-L$ distribution between two sectors, the normal 
fermionic one and the new $R_{q}$-sector. When temperature fell below 
$T<T^{\ast}$ plasma became asymmetric, $B-L \neq 0$ in both sectors.
\end{center}
\noindent

We would like to emphasize that
we want that in all the processes resulting in such an asymmetrization
of plasma B,L,(B-L)  and any other global additive quantum
numbers (or multiplicative quantum numbers like R-parity or matter parity
in supersymmetry) to be strictly conserved
both globally and locally.

Thus, after the asymmetrization the plasma remains fairly
neutral with respect to electric charge, lepton and baryon numbers,
$B-L$, etc. The only exception is obviously the fermion number
which is not conserved perturbatively. However, this might have been
not an expense at all if there were in the particle spectrum of the model
the Majorana fermions coupled to standard fermions and R-particles.

Concerning the possible mechanism of such an asymmetrization of cosmological
plasma one might expect that it might have been  provided by
CP-violating out-of-equilibrium decays of some massive Majorana fermions
(X-fermions in what follows) onto SM fermions (antifermions) and
anti-R-bosons (R-bosons) at some effective freezing-out temperature
$T^{\ast}$, $T^{\ast}> T_{EW}$, without violating any quantum number
except for fermion number,
\begin{equation}
X \rightarrow qR^{c}_{q}, q^{c}R_{q}
\label{eq:Xq}
\end{equation}
\noindent
and
\begin{equation}
X \rightarrow lR^{c}_{l}, l^{c}R_{l} .
\label{eq:Xl}
\end{equation}

\setlength{\unitlength}{1mm}
\begin{center}
\begin{picture}(135,70)(0,0)

\put(70,30){\vector(1,-1){15}}
\put(70,40){\vector(1,1){15}}
\put(15,38){\line(1,0){50}}
\put(15,32){\line(1,0){50}}
\put(65,35){\circle{10}}

\put(18,43){$X \equiv Majorana$}
\put(86,55){$q(q^{c})$}

\put(86,15){$R^{c}_{q}(R_{q})$}
\end{picture}

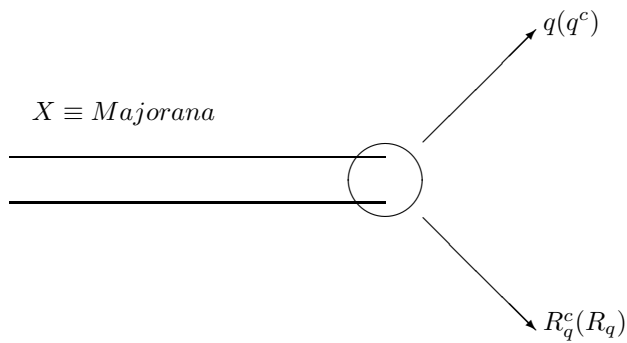
Fig. 2. A scheme of $(B-L)$-asymmetrization of plasma in charge asymmetric
decays of $X$-particles onto quarks (antiquarks) and $R_{q}
(R_{q}^{c}$-particles). The charge asymmetry might have taken place also in
decays $X \rightarrow lR^{c}_{l}, l^{c}R_{l}$.
\end{center}

\noindent
The charge asymmetry in X-decays, for example,
\begin{equation}
\Gamma (X \rightarrow qR^{c}_{q}) \equiv \Gamma_{1}
\neq \Gamma (X \rightarrow q^{c}R_{q}) \equiv \Gamma_{2}
\label{eq:Gamma}
\end{equation} 
and/or
\begin{equation}
\Gamma (X \rightarrow lR^{c}_{l}) \neq \Gamma (X \rightarrow
l^{c}R_{l}), 
\label{eq:GX}
\end{equation} 
might have arisen due to CP-noninvariance in the interference
of the tree-level diagrams and loop radiative corrections ( see Fig. 2 ),
as usual (see, e.g., the book by Kolb and Turner \cite{kolb} and the paper 
by Kuzmin and Shaposhnikov \cite{kuz4}).

In general, the amplitudes of charge-conjugated decays of X-particles
take on the form \cite{ign2} :
\begin{equation}
A(X \rightarrow a_{i}b_{i}...) = g_{i}+{\Sigma}g'_{ik}A_{ik} ,
\label{eq:Aa}
\end{equation}
\begin{equation}
A(X \rightarrow {\bar a_{i}}{\bar b_{i}}...)=
g^{\ast}_{i}+{\Sigma}g'^{\ast}_{ik}A_{ik},
\label{eq:Abar}
\end{equation}
\noindent
$g_{ik}$ being the product of corresponding coupling constants, generically 
$g_{ik} \sim f^{3}$ for one loop radiative corrections ( $f$ being the 
corresponding coupling constants in vertices), $A_{ik}$ being radiative 
corrections to the tree diagram of the decay 
taken at unity values of coupling constants. From Eqs.~\ref{eq:Aa}{eq:Abar} 
one obtains for the microscopic asymmetry $\epsilon$,
\begin{equation}
\epsilon \equiv (\Gamma - \Gamma_{CP})/{\Gamma_{tot}},    \label{eq:eps1}
\end{equation}
\begin{equation}
\epsilon = ({\frac {1}{\Gamma_{X}}})({\Gamma_{i}}B_{i}+{\Gamma_{\bar
{i}}}B_{\bar {i}})= (4{\Sigma}B_{i}Im(g^{\ast}_{i}g'_{ik})Im
A_{ik})/({\Sigma}(g_{i}g^{\ast}_{i})) ,   \label{eq:eps2}
\end{equation}
\noindent
where $\Gamma_{i}(\Gamma_{\bar {i}})$ are the partial decay widths of X into 
the channel $i({\bar {i}})$ and $B_{i}(B_{\bar {i}})$ is the baryon number 
of normal fermion (or R-particles) secondaries in the $i$-th $({\bar {i}}$-th)
channel.

The sign of the asymmetry is determined by the unknown CP-violating phase. 
One may take at the moment $\Gamma_{1}< \Gamma_{2}$.

The protection of the created charge asymmetric component of R-particles 
from disappearance due to Standard Model  
exchanges between two sectors might be achieved by the expense of 
attributing to new particles ($X$ and $R$) some new conserved 
multiplicative quantum number R.

The net $-(B-L) \neq 0$ excess in the normal left-handed SM fermionic sector 
is now becoming a subject of re-processing in the usual way by un-suppressed 
electroweak transitions in the temperature range $T^{\ast}>T>T_{EW}$ 
resulting at $T<T_{EW}$ in some baryon and lepton number asymmetries 
of plasma. The corresponding $(B-L)$ excess in the 
R-sector contained in $R_{q}$-particles remained intact by sphalerons 
and got transported to the epoch $T<T_{EW}$ just as it was created at 
$T^{\ast}$.

Having assumed that R-particles bear the 
conserved quantum number $R$ one may observe immediately that the lightest 
R-carrying particles might have survived until present epoch and  serve as a 
candidate for the cold dark matter population of the Universe.

Clearly, the number densities of excess quarks (antiquarks) and 
$R_{q}^{c}(R_{q})$-particles are equal at the production time, 
$T=T^{\ast}$, while at the end of sphaleron operating epoch at $T=T_{EW}$ 
the relation between them becomes $n_{R} \approx an_{B}$ , 
the factor $a$ lying in between 
the extreme values $a=4/3$ (if $B_{init} \neq 0, L_{init}=0$) and $a=4$ 
(if $B_{init}=0, L_{init} \neq 0$). 
At present the relation between corresponding number densities is given by
\begin{equation}
n_{R} \approx a(1-b) n_{B} , \label{eq:nR}
\end{equation}
\noindent
the factor $b$ accounting for possible depletion of asymmetric 
R-particle abundance on the way from $T=T_{EW}$ to present time. 
If the thermal charge symmetric component of R-particle content of 
plasma completely annihilated in the course of the Universe expansion 
similarly to quarks and leptons, then identifying survived relic 
R-particles with the CDM content of the Universe one arrives at the 
following estimate of their mass
\begin{equation}
m_{R} \approx (1/a(1-b))(c/d)m_{p}(\Omega_{CDM}/{\Omega_{B}}) ,
\label{eq:mR}
\end{equation}
\noindent
$m_{p}$ being proton mass and the factors $c\leq 1$ and $d\leq 1$ 
accounting for the fractions of the ${\Omega_{CDM}}$ and the total 
observed ${\Omega_{B}}$, respectively, attributed to our particular 
mechanism of the CDM and BAU generation. Clearly, it might 
be well not a unique one.

Taking $\Omega_{CDM}/{\Omega_{B}} \approx 0.7/0.05 =14$ in the mixed 
(CDM plus HDM) models one arrives 
in the extreme case $b=0, c=1, d=1$ to the estimate
\begin{equation}
m_{R} \approx (14/a) GeV. \label{eq:14/a}
\end{equation}

What is very important is the following. The ratios  
of the produced in such a way cosmological baryon 
excess and CDM content seem to be insensitive to the 
character ( 1st or 2nd order) of the electroweak phase transition, 
in contrast to the common case when efforts of solving the cosmological 
baryon excess problem within the framework of the Standard Model itself.

Thus,  the essence of our scenario of a possible common genesis 
of the BAU and the CDM in the Universe is a preparation of a state 
of plasma with $B-L \neq 0$ in the fermionic sector of the 
SM and $-(B-L)$ in the new particle sector R, the standard fermions 
being involved in sphaleron-mediated $B,L$-non-conserving processes 
while the baryon or 
lepton number bearing R-particles are sphaleron-blind. No violation 
of $B$ and/or $L$ other than that provided by sphalerons is necessary. 
Subsequent sphaleron re-processing of the $B-L$ excess in SM sector gives 
rise to the BAU and lightest stable massive R-particles contribute 
to the CDM.

Masses of X-particles necessary to provide 
generation of the observed BAU, 
\begin{equation}
\Delta \equiv n_{B}/n_{\gamma} \sim 10^{-10}. 
\label{eq:Del}
\end{equation}
might be found from consideration of the process of generation of the 
asymmetry and its washing-out \cite{ks1}. The resulting macroscopic 
asymmetry in the out-of-equilibrium decay mechanism is known to be  
given generically by \cite {ks1}
\begin{equation}
\Delta \sim (45{\zeta} (3)/4{\pi^{4}}N){\Sigma}N^{i}{\epsilon_{i}}S_{i} ,
\label{eq:Delta}
\end{equation}
\noindent
where $N$ is the effective number of degrees of freedom of massless at 
the given temperature $T$ particles, $\zeta$ is the Riemann function, 
$\epsilon$ is the microscopic asymmetry in the decay of a parent 
particle, and $S$ is the macroscopic suppression factor \cite {ks1} 
arising due to baryon number dissipation in decay and inverse decay 
processes as well as scattering of the product particles. It is generically
\begin{equation}
S \leq 10^{-2} .                         \label{eq:S}
\end{equation}
\noindent

One may conjecture that the asymmetry $\epsilon$ might be small enough in 
order to be able to explain the observed baryon asymmetry of the Universe. 
This might be just the case, indeed. However, even in this case the proposed 
mechanism of asymmetrization of cosmological plasma may provide the origin of
a charge asymmetric CDM component of the Universe. This latter might be 
electrically neutral as well as (negatively) charged. This case is obviously 
of a special interest.  

\section{Realizations of the Scenario in the Framework of  Supersymmetric
Models.}

Let us examine in this respect a supersymmetric extension of the
Standard Model, for example, let us consider the Minimal Supersymmetric
Standard Model
(MSSM) in order to clarify its resources. One finds
that there seems to be quite enough room even within this simplest
supersymmetric model for a realization of
the scheme, at least in a sense of some asymmetrization of plasma.
Indeed, our R-particles could be nothing but sfermions which bear
baryon or lepton number. However, they are the Lorentz scalars and 
therefore are not affected by sphalerons. 
Further, there are Majorana fermions in the supersector, 
just gauginos, ${\tilde{B}^0}$ (bino), ${\tilde{W}^0_3}$ (wino) 
and ${\tilde{g}}$ (gluino) before $SU(2)_L \times U(1)$ breaking, so
\begin{equation}
X \equiv {\tilde {B}^0}, ( {\tilde {W}^0_3},  {\tilde {g}}).
\label{eq:XB}
\end{equation}
\noindent
After $SU(2)\times U(1)$ breaking at electroweak scale, $T \sim M_{W}$, 
these become
\begin{equation}
\tilde{\gamma} , \tilde{Z^{0}}, \tilde{g}. 
\label{eq:phot}
\end{equation}
in mixtures. There are also $\tilde{H}_{1}$ and $\tilde{H}_{2}$. 
In supergravity case it might be also that it is just gravitino which 
plays a role of a parent particle in baryogenesis and CDM genesis,   
\begin{equation}
X \equiv {\tilde {G}}, 
\label{eq:XG}
\end{equation}
where ${\tilde {G}}$ denotes gravitino.

As an example, we shall consider just bino ${\tilde{B}}^{0}$  
decays, the cases of ${\tilde{W^0_3}}$, ${\tilde{g}}$ or ${\tilde {G}}$ 
being quite similar.

It goes without saying that these gauginos are to be massive at 
$T>T^{\ast}$,
\begin{equation}
m_{\tilde {B^0}} > T^{\ast} , \label{eq:mB}
\end{equation}
\noindent
i.e. we assume here that supersymmetry is broken at scales higher 
than $T^{\ast}$.

It is clear that there might have taken place two extreme cases, namely, 
the maximal 
$B-L$ asymmetry in the normal sector being due to leptonic deacays of 
X-particles, 
or due to decays of X onto squarks {anti-squarks) and $R_{q}(R_{q}^{c})$, 
depending on the amount of CP-violation, i.e.  coupling constants and 
CP-angles. This does not make any principal difference but two cases 
deserve detailed analysis. We shall restrict ourselves for demonstration 
purposes by the quite short description of the case 
when all the $(B-L)$- asymmetry comes from decays of X into baryonic 
sector (i.e. $B_{initial} \neq 0, L_{initial}=0$, see below.) 
Clearly, this is an oversimplifying description of what might have 
occurred. In facr, both asymmetries took place simultaneously and 
are to be taken into account.
    
By obvious reasons of largest couplings to Higgs bosons of 
top quarks and top-squarks, one may expect that this will result in the 
largest radiative corrections to the tree level diagrams of 
bino decays and therefore in largest asymmetry in just these decays. 
We shall therefore be interested mainly just in the processes like 
\begin{equation}
{\tilde{B}}^{0} \rightarrow t{\tilde{t}}^{c},t^{c}{\tilde{t}}.
\label{eq:Bt}
\end{equation}
\noindent

All other decay channels of all the gauginos onto quarks of 1st and 2nd 
generations, 
\begin{equation}  
{\tilde{B}}^{0} \rightarrow q{\tilde{q}}^{c},q^{c}{\tilde{q}} , 
q \equiv u,d,c,s  \label{eq:Bq} 
\end{equation}
or lepton decays,
\begin{equation} 
{\tilde{B}}^{0} \rightarrow l{\tilde{l}}^{c},l^{c}{\tilde{l}} 
\label{eq:Bl} 
\end{equation}
might be expected to be less efficient. We are not going though to 
overestimate the validity of such kind of arguments. This is simply an 
example of our line of reasoning. As soon as the model is specified 
one needs not any further assumptions. 

Clearly, one has to assume 
\begin{equation}
m_{\tilde{B}}^{0}>m_{\tilde{t}}. \label{eq:mBmt}
\end{equation}
\noindent

In fact, as one can see, we have to require masses of all gauginos  
to be bigger than those of all the sfermions, 
\begin{equation}
m_{gaugino}> m_{sfermion}. \label{eq:gsf}
\end{equation}

This is not a commonly taken point of view. However, it might be not 
quite stupid while taking into account the renormalization group equation
of evolution of coupling constants with proper values of $m_{0}$ and 
$m_{1/2}$. 

We emphasize that 
no violation of $R$-parity or $B$ and/or $L$ is necessary in these processes.

As soon as one does not assume any R-parity violation, neither 
explicit nor spontaneous, the lightest sparticles (LSP) are stable, 
as usually.

What happened to the originated at $T=T^{\ast}$ charge asymmetric spartner  
component depends upon which of all sparticles is the LSP. There is 
${\it a priori}$ a number of possibilities. However, according to the 
very idea of the scenario, one has to require that after the temperature 
has fallen down to $T=T^{\ast}$ any $B$ and $L$ transfer from one sector 
to another was to be effectively switched off. Therefore, not only 
gauginos but higgsinos as well are to be heavier than sfermions, 
\begin{equation}
m_{\tilde {H}}>m_{\tilde {f}}, {\tilde{H}} \equiv
{\tilde{H}}_{1},{\tilde{H}}_{2}. 
\label{eq:H}
\end{equation}
\noindent
Otherwise there might have taken place too fast decays of squarks into 
ordinary quarks, 
\begin{equation} 
{\tilde{q}} \rightarrow q{\tilde{H}},  \label{eq:qH}
\end{equation}
before sphalerons got frozen-out 
of equilibrium. Such decays would just mean some returning of baryon 
number back to the normal sector. Choosing between two possibilities, 
a squark or a slepton being the LSP, one definitely prefers by 
several reasons the latter one. Therefore, the squark excess after 
$T=T_{EW}$ is to be converted into sleptons. This might have been 
fairly naturally provided by  squark decays like ( see Fig. 3 )

\begin{equation}
{\tilde{t}} \rightarrow tl{\tilde{l}}^{c},tl^{c}{\tilde{l}} .
\label{eq:tt}
\end{equation}

Thus, there takes place a quite remarkable total return of 
the  'temporarily loaned' baryon
number from the supersector to the normal SM quark sector. However, it
does not anymore compensate exactly the $B$ excess in the normal
sector since the latter has suffered from partial sphaleron
re-processing. 

The resulting output overall baryon excess (contained
exclusively in the normal quark sector) is positive, $B_{final}>0$,
and is given by
\begin{equation}
B_{final} \approx (1/4)B_{initial}, \label{eq:Bfin}
\end{equation}

This completes the story.

\setlength{\unitlength}{1mm}
\begin{center}
\begin{picture}(130,60)(0,0)

\put(30,43){${\tilde t}$}
\put(15,40){\line(1,0){5}}
\put(23,40){\line(1,0){5}}
\put(31,40){\line(1,0){5}}
\put(39,40){\line(1,0){5}}
\put(46,40){\line(1,0){40}}
\put(89,40){$t$}
\put(50,25){${\tilde B^o}$}
\put(46,40){\line(1,-1){16}}
\put(62,24){\line(1,-1){4}}
\put(68,18){\line(1,-1){4}}
\put(74,12){\line(1,-1){4}}
\put(80,06){\line(1,-1){4}}

\put(89,24){$l^{c} (l)$}
\put(89,0){${\tilde l} ({\tilde l^c})$}
\put(62,24){\line(1,0){24}}

\put(73,19){\oval(3,10)}
\end{picture}
\end{center}
\vspace{1cm}
\begin{center}
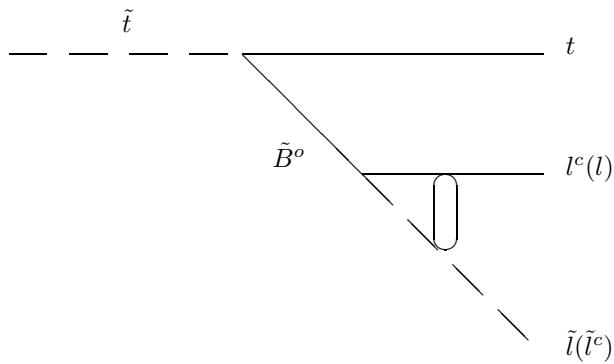
Fig. 3. A diagram showing the return of the  baryon number excess contained 
in supersymmetric sector to the normal quark sector of the Standard Model 
and creation of the final
CDM content of the Universe in the form of sleptons (antisleptons).
\end{center}
\vspace{1cm}
\noindent

One can easily see that the freezing-out temperature of ${\tilde {t}}$ 
is to be lower than $T_{EW}$ ( i.e. ${\tilde {t}}$ should disappear from 
plasma after temperature had fallen down $T_{EW}$) 
in order not to return the baryon excess contained in the supersector to 
the normal quark sector too early. This means that ${\tilde {t}}$ must 
be light enough,
\begin{equation}
m_{\tilde{t}} \leq 20 T_{EW} \approx 2 TeV ,  \label{eq:tT}
\end{equation}
\noindent
and there are sleptons in the spectrum which are light enough,  
\begin{equation}
m_{\tilde{l}}< ((1/2)m_{\tilde{t}}-m_{t}) \leq 1 TeV . \label{eq:lt}
\end{equation}

\subsection{Charge Symmetric Slepton Component of CDM.}

If decays of ${\tilde{t}}$, Eq.~\ref{eq:tt}, are charge symmetric and 
sleptons are lightest (stable) superparticles then this 
will result in creation of charge symmetric (slepton) cold dark 
matter component of the Universe with their number density twice as large 
as the ${\tilde{t}}$'s. This will result in the very low estimate of their 
mass, Eq.~\ref{eq:14/a}, $m_{\tilde{l}} \sim 2$ GeV (see Section 4). 

This is by no means acceptable for any left-handed sleptons due to 
corresponding contribution to the total $Z^0$-width.

Therefore, the charge symmetric component of these deacays can not represent
the CDM. Having originated from these decays, it  effectively 
disappears from plasma due to subsequent annihilation.

\subsection{Charge Asymmetric Slepton Component of CDM.}

The very interesting point is however the following. The slepton-antislepton 
component originated from decays of squark excess might have had again a tiny 
charge asymmetry ${\delta}$ due to radiative corrections to 
the (virtual) bino vertex ${{l}^{c}}{\tilde{\l}}{\tilde{B^0}}$.  
The most promising asymmetric decay channels are presumably the ones with 
${\nu_{\tau}},{\tilde \nu}_{\tau}^{c}$ due to the largest Higgs couplings, 
\begin{equation}
{\tilde{t}} \rightarrow t{\tilde \nu_{\tau}}{\nu_{\tau}^c},
t{\tilde{\nu_{\tau}^c}}{\nu_\tau},
\label{eq:tnu}
\end{equation}
and decays with charged sleptons ${\tau}{\tilde\tau}^{c}$ in the final state
\begin{equation}
{\tilde {t}} \rightarrow t{\tau}{\tilde \tau^c}, t{\tau^c}{\tilde \tau}.
\label{eq:ttau}
\end{equation}

One may expect that this charge asymmetry, $\delta$, might be 
presumably of order 
${\delta} \leq 10^{-6}$. Hence, the relation between the excess 
baryon and asymmetric slepton number densities becomes
\begin{equation}
n_{\tilde{l}} \sim 4 \delta n_{B} . \label{eq:delta}
\end{equation}

It is worth noting that this would-be CDM asymmetric slepton component 
has a non-thermal momentum spectrum.

Neglecting the depletion of slepton number density due to two slepton 
pair-annihilation processes after temperature has dropted below 
$T_{EW}$
\begin{equation}
{\tilde{l}}{\tilde{l}} \rightarrow ll  \label{eq:ll}
\end{equation}
\noindent
which is possible because of $R$-parity being a multiplicative quantum 
number one obtains an estimate of the possible CDM content 
due to this asymmetric component  
using Eq.~\ref{eq:Bfin}, Eq.~\ref{eq:delta}:
\begin{equation}
\Omega_{CDM}/{\Omega_{B}} \sim 4. 10^{-3}, \label{eq:Om}
\end{equation}  
in the case of all the observed BAU,  $\Omega_{B} \approx 0.1$, being due to
our mechanism,
$\delta \leq 10^{-6}$ and $m_{\tilde{l}} \leq 1$ TeV.

Yet two possibilities are now in turn in this charge asymmetric dark 
matter scenario, namely, the LSP being either 1) the left-handed sneutrino, 
or 2) the charged slepton. None of these seems to be excluded 
${\it a priori}$.

{1. \it Neutral $SU(2)_{L}$-doublet slepton as LSP}. 

If just the ($SU(2)_{L}$-doublet) sneutrino is the LSP then 
the overall output of the charge asymmetric CDM scenario is quite 
similar to the commonly used one except for the smallness of the 
corresponding CDM content, ${\Omega_{CDM}}/{\Omega_{B}} \sim 4.10^{-3}$, 
Eq.~\ref{eq:Om}, which being natural does not pretend nevertheless to explain 
all the CDM content of the Universe.  

The estimate $m_{\tilde{\nu}} \leq 1$ TeV does not come into 
contradiction with any known constraints on sneutrino mass. 
The counting rate in experiments devoted to direct searches of the 
flux of weakly interacting massive particles (WIMP) from the 
galactic halo is smaller than is usually expected. 
 
{2. \it Charged Slepton as LSP.} 

Quite a different and exciting possibility might have been realized 
if just a charged slepton is the LSP. The possibility that stable charged 
particles, in particular, sleptons might constitute the CDM 
was analyzed in the paper by De Rujula {\it et al} \cite{DeR} 
(where these particles were called champs). 
An exciting story of the evolution of the relic champs content 
in the Universe was pictured out and it was argued that the case of champs 
might be not excluded by current observations. We would like to add 
few remarks.

In our case, the CDM is assumed to be charge asymmetric and 
consists of negatively charged sleptons. 
It is interesting to note that our estimate of slepton mass 
Eq.~\ref{eq:lt}, $m_{\tilde{l}} \leq 1$ TeV, does not stay 
catastrophically apart from the window of allowed champ 
masses $10 - 1000$ TeV obtained by De Rujula {\it et al} \cite{DeR} from 
different arguments. Thus, we would consider our negative sleptons 
(asymmetric component) 
as a reasonably good candidate for champs.

Starting with the time of origination from the excess squark decay at 
$T<T_{EW}$ and down to the temperature of order 
$T \sim few$ $hundreds$ KeV nothing essential happened to 
${\tilde{l}}$ excess. Drastic phenomena occurred \cite{DeR} after 
$T$ had fallen down to $T \sim$  few hundreds KeV when the primordial 
nucleosynthesis began to proceed. Now ${\tilde{l}}$ came into play. 
They took part in nucleosynthesis processes catalyzing them to some 
extent as well as got starting to proceed through complicated kinetics 
of recombination processes.They were getting 'dressed' by protons and 
$\alpha$'s 
and forming atoms like $({\tilde{l}}p)$ (superhydrogen in what 
follows) with binding energy 
\begin{equation}
E_{b} \approx 25 KeV, \label{eq:Eb}
\end{equation}
\noindent  
as well 
as ions like $({\tilde{l}}\alpha)$, ($E_{b} \approx 311$ KeV \cite{DeR}), 
and atoms of superhelium $({\tilde{l}}{\tilde{l}}\alpha)$, with the 
binding energy of about 800 KeV, etc. 
According to De Rujula {\it et al} \cite {DeR} 'negative champs 
overwhelmingly bind to protons to pose as super-heavy neutrons' called 
in \cite{DeR} neutrachamps. In our case a neutrachamp is $({\tilde{l}}p)$. 
For definiteness, let us take selectron, ${\tilde e}$,  as the LSP. 

Atoms $({\tilde{e}}{\tilde{e}}\alpha)$ in which two ${\tilde{e}}$
are getting dressed by $\alpha$-particle are in any case unstable and have 
short lifetimes in cosmological scales due to pair-annihilation process of 
two ${\tilde{e}}$ into ordinary leptons. 

After finishing the ${\tilde{e}}$ recombination period and formation of
superhydrogen atoms $({\tilde{e}}p)$ and then the recombination period 
for (normal) hydrogen and helium,  
the next important stage in the evolution is met right at 
formation of galaxies and clusters of galaxies. 
The gas of superhydrogen will presumably share the fate of all other 
gases at this stage, so it will be as abundant in the galactic matter 
at this time as it does in cosmological plasma.

Further, of all the neutral gases (hydrogen,
helium, superhydrogen, etc) the gas of neutral superhydrogen
is the most collisionless because of compactness of the atom,
the mean size of it  being $r \sim 2.10^{-12}$ cm. 

Therefore, one might expect that at the next important stage of the 
evolution, namely, star formation inside galaxies,  superhydrogen
atoms were not effectively involved in contraction processes due to  
lack of tisssssssme and were left
not clustered inside the Galaxy constituting a widely distributed CDM
content with velocities $v \sim 10^{-3}$ and the local density somewhat 
about 
\begin{equation}
\rho_{\tilde{e}p} \sim 4.10^{-3} \rho_{local} \sim 10^{-3} GeV/cm^{3} ,  
\label{eq:rho}
\end{equation}
according to Eq.~\ref{eq:Om}. Here $\rho_{local} \approx 0.3$ $GeV/cm^{3}$ 
is usually taken 
local dark matter density. 
The number density of superhydrogen atoms will be then
\begin{equation}
n_{\tilde{e}p} = \rho_{(\tilde{e}p)}/m_{\tilde{e}} \sim 10^{-6} cm^{-3} 
\label{eq:ne}
\end{equation}
\noindent  
if the mass of $({\tilde{e}})$ is about 1 TeV, Eq.~\ref{eq:Om}. 
Hence, the local flux intensity of our 
superhydrogen atoms in the space might in be expected to be of order 
\begin{equation}
F_{(\tilde{e}p)} \sim 30 cm^{-2}s^{-1} .  \label{eq:Fe}
\end{equation}   

If so, there would be quite small 
${\it primordial}$ abundance of superhydrogen inside the Sun and the Earth. 
These bodies got to start absorbing the flux of superhydrogen from the space
as soon as would-be-star clouds became condensed enough.

The total amount of $({\tilde{e}p})$ accumulated by
the Earth through
all the terrestrial history as condensed body might then be about 
$10^{36}$, their
average (over the Earth) relative abundance being about 
\begin{equation} 
n_{\tilde{e}p}/n_{nucl} \sim 10^{-15} . \label{eq:nn} 
\end{equation}
\noindent
This is quite an admixture of wild isotopes to normal element 
abundances even on average! 

Note that there takes place a quite remarkable phenomenon of
fast enough changing by ${\tilde{e}}$'s their host nuclei from protons 
in superhydrogen
to nuclei with larger atomic numbers. The energy release in this process is
about $ E \sim 25Z^{2}A$ KeV, i.e., for example, in the case of iron $^{56}Fe$ 
\begin{equation}
({\tilde{e}}p)+{^{56}}Fe \rightarrow ({\tilde{e}}^{56}Fe)+p +{\pi}'s + {\gamma}'s 
\label{eq:eFe}
\end{equation}
\noindent
it is about $E \sim 800$ MeV while in case of oxygen it is about 1 MeV. 
Therefore, all the superhydrogen atoms falling down the Earth's atmosphere 
are captured by nuclei of nitrogen, oxygen, carbon, etc. 
Clearly, this will result in emission of quite characteristic hard Roentgen 
$\gamma$'s from the top of the atmosphere with well determined energies. 
Obviously, this radiation is to be searched for. 
     
The situation is even more exciting in case of the Moon.  
Here all the accumulated amount of ${\tilde{e}}$ transferred from 
superhydrogen atoms to heavier nuclei is contained in a quite thin 
layer of the Moon ground just near the surface, so the relative abundance 
of wild heavy isotopes should be larger by orders of magnitude than 
Eq.~\ref{eq:nn}. It seems therefore that search of relic selectron  
abundance might be most promising by analysis of chemical content of  
samples of the Moon ground. Methods of laser spectroscopy providing 
sensitivity to contamination up to $10^{-16}$ might be well adequate. 

Being binded to protons very strongly, 
$E_{b} = 25$ KeV, selectrons 
are not probably taking part in acceleration processes resulting in 
cosmic ray production in objects like supernovae, since temperatures 
are hardly high enough for ionization of superhydrogen atoms. However, 
nevertheless there should be some flux of bare negative selectrons in 
cosmic rays due to interaction of primary cosmic rays with the 
superhydrogen gas during their travel for $\sim 20$ million years 
inside the Galaxy. Clearly, the flux of bare selectrons  
from the space will be superpenetrative even in comparison with muons 
produced in the atmosphere because of selectrons' larger mass and stability. 
They might be looked for very deep underground.

The very intriguing at first sight issue,  why the flux
$F_{(\tilde{e}p)} \sim 30 cm^{-2}s^{-1}$ of superhydrogen atoms
from the outer space was not observed in experiments devoted to
the CDM searches, is quite easy to explain.  
The flux of superhydrogen atoms
is expected to be about $10^{3}$ times less intensive than usually 
expected one in case of WIMPS with masses of order 100 GeV but 
the cross-section of interaction with nuclei is much bigger since 
they are interacting stronly and electro-magnetically rather than 
weakly. So, the effect per ingoing particle is orders of magnitude bigger 
than in the case of WIMP's.

However, the main possible reason for non-observation of superhydrogen 
atoms might is related to absorption of
superhydrogen atoms en route to detectors.
(One has to take into account that being aimed to
look for rare events of
nuclei getting small recoils due to weakly interacting
particles of CDM these experiments are  being carried
out usually in underground laboratories. 
One has presumably to explore small or shallow 
depths, not to say satellites, where the effect itself would be bigger by 
the ratio of cross-sections, i.e. by many orders of magnitude since 
superhydrogen atoms are interacting with matter electro-magnetically and 
strongly and do not penetrate too far deep.)

\section{MSSM plus ${\nu}_{R}$ and ${\tilde{\nu}}_{R}$.}

Until now we considered the case of the supersymmetrized version of the 
Standard Model without right-handed neutrinos and sneutrinos. If 
one takes into account possible existence of these particles, then 
one may arrive at the possible explanation of ${\it all}$ the  
baryon excess and ${\it all}$ the CDM content in the Universe, 
$\Omega_{CDM} \sim 0.7$, as being produced simultaneously 
according to our mechanism.      

In this case the number densities of $({\tilde{\nu}}_{R}$ and 
${\tilde{\nu}}^{c})$ are equal and each is about 
\begin{equation}
n({\tilde{\nu}}_{R}) \approx 4n_{B} , 
\label{eq:nuB}
\end{equation}
so, the mass of each of these species is
\begin{equation}
m({\tilde{\nu}}_{R}) \approx 1.8 GeV .
\label{eq:nuR}
\end{equation}
\noindent

Note that in this case one arrives not at the constraint
on the mass but just at the prediction of the concrete value of it 
according to Eq.~\ref{eq:14/a}. 
The uncertainty in Eq.~\ref{eq:nuR} is only related with the ratio 
$(\Omega_{CDM}/\Omega_{B})$. 
It is a very striking and straitforward consequence of the very concept. 

It does not however seem to be quite an absurd from the point of view 
of renormalization group evolution of coupling constants with proper 
values of $m_0$ and $m_{1/2}$.
 
We have to note by the way that with this estimate of ${\tilde\nu}_{R}$ 
mass one should care 
about the see-saw mass for neutrino, lepton number violation due to 
Majorana neutrino mass, and so on. We will consider all this stuff in the  
forthcoming paper \cite{kuz5}.

Being $SU(2)_L$-singlets
they do not suffer any significant depletion of their number densities due to
annihilation.

The contribution of ${\tilde{\nu}}_{R}$ and/or ${\nu}_{R}$ 
to $Z^0$ total width (see Fig. 4) 
might have been dangerous in the case of large ${\tilde{\nu}}_{R}
{\tilde{\nu}}_{L}$ and ${\nu}_{R}{\nu_{L}}$ mixing. 
Fortunately, such mixing is small enough and is 
not excluded by measurements of the total $Z^{0}$-width. 

Two obvious circumstances make ${\tilde\nu}_{R}$ as a candidate for CDM 
very hard to observe. 
 
1. The smallness of the
${\tilde{nu}}_{R}$ mass, Eq.~\ref{eq:nuR}, will lead to much smaller 
nuclei recoil energies, $E_{recoil} \sim 1$ KeV in comparison with 
usually expected $E_{recoil} \sim 50- 100$ KeV in underground experiments 
devoted to searches for 
weak interacting particles with masses of order $100$ GeV.
Therefore, the signal from light ${\tilde\nu}_{R}$ scattering off nuclei will 
require very low thresholds. 

2. In addition, the very rate of scatterings of ${\tilde\nu}_{R}$  
should be very low because ${\tilde\nu}_{R}$ is neutral $SU(2)_{L}$-singlet.
   
\setlength{\unitlength}{1mm}
\begin{center}
\begin{picture}(130,50)(0,0)

\put(30,20){\oval(4,4)[t]}   
\put(34,20){\oval(4,4)[b]}

\put(38,20){\oval(4,4)[t]}   
\put(42,20){\oval(4,4)[b]}
\put(46,20){\oval(4,4)[t]}   
\put(50,20){\oval(4,4)[b]}

\put(40,25){$Z^o$}           
\put(53,20){\line(-1,0){1}}

\put(53,20){\line(1,1){4}}   
\put(59,26){\line(1,1){4}}
\put(65,32){\line(1,1){4}}   
\put(71,38){\line(1,1){4}}

\put(54,27){${\tilde \nu_L}$}
\put(72,35){${\tilde \nu_R}$}
\put(54,12){${\tilde \nu_L^c}$}
\put(72,4){${\tilde \nu_R^c}$}

\put(64,31){\circle*{2}}

\put(53,20){\line(1,-1){4}}
\put(59,14){\line(1,-1){4}}
\put(65,8){\line(1,-1){4}}
\put(71,2){\line(1,-1){4}}

\put(64,9){\circle*{2}} 
\end{picture}
\end{center}
\vspace{20mm} 

\begin{center}
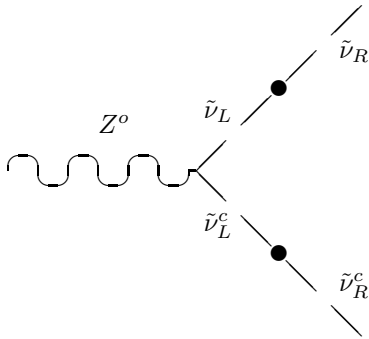
Fig. 4. A diagram of decay $Z^{0} \rightarrow {\tilde\nu_R \tilde\nu^c_{R}}$ 
( or $Z^0 \rightarrow {\tilde \nu_L \nu_R^{c}}$ 
if $m_{\tilde \nu_{L}} < m_{Z} -m_{\tilde\nu_{R}}$; 
in the latter case there is 
only one $({\tilde\nu_{L}},{\tilde\nu_{R}})$ mixing insertion). All the same 
refers to ${\nu}_{R}$ and ${\nu}_{L}$.
\end{center} 
\vspace{10mm}
\noindent
The partial width $Z^{0} \rightarrow {{\tilde\nu}_R} {\tilde\nu}_{R}^c$ 
is proportional to ${\sin}^{4}{\theta}$, ${\theta}$ being the 
${\tilde{\nu}}_{R}{\tilde{\nu}}_{L}$ mixing angle. The mixing is due to 
the $SU(2)_{L} \times U(1)$ breaking. The $\theta$ might be 
expressed in terms of coupling constants and the Higgs' boson 
vacuum expectation value. 

If ${\tilde\nu}_{R}$ is the lightest sparticle indeed, then we predict that 
there will be quite long-living spartners in the spectrum. 
This follows obviously from the fact of necessary mixing of left-handed 
and right-handed components of sneutrinos in this case which is small. 
Of particular interest is the prediction of existence of charged long-living 
sleptons. This should be 
taken into account in the searches of sparticles in accelerator experiments 
and, possibly, in deep underground cosmic ray experiments. This is by itself 
a very striking consequence of the scenario.

\section{Conclusions.}

In this paper we presented the new concept of a possible origin of the  
simultaneous production of the baryon excess and cold dark matter in the 
Universe. The basic expense is the assumption on the existence in Nature 
of particles ( R-particles ) which bear baryon or lepton numbers 
but are sphaleron-blind. As an example, we considered the case of 
R-particles being Lorentz scalars using for illustrative purposes 
supersymmetric models with their generic particle content. 

It is interesting that generically any version of our scenario of 
simultaneous 
production of the cosmological baryon excess and cold dark matter 
in the Universe leads presumably to the prediction of the Cold Dark 
Matter content 
in the form of  ${\it superweak}$ interacting and hard-to-observe 
in direct CDM search experiments very light 
particles with masses of about 2 GeV. 

In the case of supersymmetric realization of the basic idea 
the CDM is nothing but right-handed sneutrinos with 
$m_{{\nu}_{R}} \approx 2$ GeV.   
 
The very interesting version of the scenario is 
the one with the charge asymmetric CDM content, more specifically 
with charged sleptons as the LSP which got dressed by protons forming 
compact stable neutral superhydrogen atoms. The estimated masses are 
$m_{\tilde{l}} \leq 1$ TeV. These are not abundant very much, however,
it is worthwhile to look for them.

\section*{Acknowlegements.}

I am grateful to A. Bottino, D. Cline, J. Ellis, A.Yu. Ignatiev, 
H.V. Klapdor-Kleingrothaus, N.V. Krasnikov, V.M. Lobashev, 
R.N. Mohapatra,  L.B. Okun, J. Pati, V.A. Rubakov, S. Ruby, G. Senjanovic,  
M.E. Shaposhnikov, A.Yu. Smirnov, G. Steigman, L. Stodolsky, 
P.G. Tinyakov, I.I. Tkachev, and V.I. Zakharov 
for helpful discussions. I am grateful to E.Kh. Akhmedov, D. Tommasini and 
especially J.F.W. Valle for stimulating discussions at the beginning of this 
work.  I am thankful very much to 
L. Stodolsky for his extreme hospitality extended to me during my stay at 
Max-Planck Institut fuer Physik, Muenchen, and H.V. Klapdor-Kleingrothaus 
for hospitality at Max-Planck Institut fuer Kernphysik, Heidelberg.


\section*{References}

\begin{thebibliography} {99}

\bibitem{sakh}
A.D. Sakharov, Pis'ma ZhETF, {\bf 5} (1967) 32
(JETP Letters {\bf 5} (1967) 24.)

\bibitem{kuz1}
V.A. Kuzmin, Pis'ma ZhETF, {\bf 12} (1970) 335.

\bibitem{ign}
A.Yu. Ignatiev, N.V. Krasnikov, V.A. Kuzmin, and A.N. Tavkhelidze, Proc.
Int. Conf. Neutrino-77, vol. 2 (Nauka Publ., Moscow, 1978) p.293; Phys. Lett.
{\bf 76B}(1978) 436;\\ 
M. Yoshimura, Phys. Rev. Lett. {\bf 41}(1978) 281; {\bf (E)42}(1979) 476;\\ 
S. Weinberg, Phys. rev. lett. {\bf 42}(1979) 850;\\
A.Yu. Ignatiev, V.A. Kuzmin, and M.E. Shaposhnikov, Phys. Lett. {\bf 87B}
(1979) 114.\\ 
For a review see, e.g.,\\ 
P. Langacker, Phys. Rep. {\bf 72}(1981) 185,  \\
A.D. Dolgov, Phys. Rep. {\bf 222} (1992) 309, and most recently \\
V.A. Rubakov and M.E. Shaposhnikov, Sov. Phys. Usp. {\bf 166} (1996) 493.
 
\bibitem{krs}
V.A. Kuzmin, V.A. Rubakov, and M.E. Shaposhnikov, Phys. Lett. {\bf B155}
(1985) 36.

\bibitem{fuku}
M. Fukugita and T. Yanagida, Phys. Lett. {\bf B174}(1986) 45.
\bibitem{lan}
P. Langacker, R. Peccei, and T. Yanagida, Mod. Phys. Lett. {\bf A 1}
(1986) 541.

\bibitem{shap}
M.E. Shaposhnikov, JETP Lett. {\bf 44}(1986) 465; Nucl. Phys. {\bf B287}
(1987) 757; {\bf B299}(1988) 797.

\bibitem{prim}
J. Primack, talk at this Workshop.

\bibitem{barr}
S.M. Barr, R.S. Chivukula, and E. Farhi, Phys. Lett. {\bf B 241}(1990) 387;
S.M. Barr, Phys. Rev. {\bf D44}(1991) 3062.

\bibitem{kap}
D.B. Kaplan, Phys. Rev. Lett. {\bf 68}(1992) 741.

\bibitem{kuz3}
V.A. Kuzmin, M.E. Shaposhnikov, and I.I. Tkachev, Phys. Rev. {\bf D45}
(1992) 466.

\bibitem{abj}
S. Adler, Phys. Rev. {\bf 177}(1969) 2426;\\
J.S. Bell and R. Jackiw, Nuovo Cimento {\bf 51} (1969) 47.\\
W.A. Bardeen, Phys. Rev. {\bf 184} (1969) 1841.
\bibitem{cdj}
R. Jackiw and C. Rebbi, Phys. Rev. Lett. {\bf 37} (1976) 172;\\
C.G. Callan, D.F. Dashen, and D. Gross, Phys. Lett. {\bf 63B} (1976) 374. 
\bibitem{kman}
F.R. Klinkhamer and M.S. Manton, Phys. Rev. {\bf D30}(1984) 2212.
\bibitem{christ}
N.H. Christ, Phys. Rev. Phys. Rev. {\bf D21} (1980) 1591.
\bibitem{bel}
A. Belavin, A. Polyakov, A. Schwarz, and Yu. Tyupkin, Phys. Lett. {58B}
(1975) 85.
\bibitem{tho}
G. 't Hooft, Phys. Rev. Lett. {\bf 37} (1076) 8.
\bibitem{kir}
D.A. Kirzhnits, JETP Lett. {\bf 153} (1972) 529;\\
D.A. Kirzhnits and A.D. Linde, Phys. Lett. {\bf 42B} (1972) 471.
\bibitem{armc}
P. Arnold and L. McLerran, Phys. Rev. {\bf D36} (1987) 581.
\bibitem{kolb}
E.W. Kolb and M.S. Turner, The Early Universe, Addison-Wesley Publ. Comp.,
(1990).
\bibitem{kuz4}
V.A. Kuzmin and M.E. Shaposhnikov, preprint INR-P-0213 (1981).

\bibitem{ign2}
A.Yu. Ignatiev, V.A. Kuzmin, and M.E. Shaposhnikov, Pis'ma ZhETF, {\bf 30}
(1979) 726.

\bibitem{ellis}
J. Ellis, D.V. Nanopoulos, and K.A. Olive, CERN preprint TH.6721/92 (1992).

\bibitem{DeR}
A. De Rujula, S.L. Glashow, and U. Sarid, Nucl. Phys. {\bf B333}(1990) 173.

\bibitem{camp}
B.A. Campbell, S. Davidson, J. Ellis, and K.A. Olive,
preprint CERN-TH-6642/92, (1992).

\bibitem{ks1}
V.A. Kuzmin, and M.E. Shaposhnikov, Phys. Lett. {\bf 105B} (1981) 163;
preprint INR, P-0190 (1981).

\bibitem{kuz5}
V.A. Kuzmin, work in progress.

\end{thebibliography}
\end{document}